\documentclass[prb,reprint,twocolumn]{revtex4-1} 

\usepackage{amsmath}  
\usepackage{amsfonts} 
\usepackage{graphicx} 

\begin{document}


\title{Simulating the photometric study of pulsating white dwarf stars \\
       in the physics laboratory}

\author{Paul Chote}
\email{paul.chote@vuw.ac.nz} 

\author{Denis J. Sullivan}
\email{denis.sullivan@vuw.ac.nz} 


\affiliation{School of Chemical and Physical Sciences, Victoria University of
Wellington, PO Box 600 Wellington 6012, New Zealand}

\date{\today}

\begin{abstract}
We have designed a realistic simulation of astronomical observing
using a relatively low-cost commercial CCD camera and a
microcontroller-based circuit that drives LEDs inside a light-tight
box with time-varying intensities.  As part of a laboratory
experiment, students can acquire sequences of images using the camera,
and then perform data analysis using a language such as MATLAB or
Python to: (a) extract the intensity of the imaged LEDs, (b) perform
basic calibrations on the time-series data, and (c) convert their data
into the frequency domain where they can then identify the frequency
structure.  The primary focus is on studying light curves produced by
the pulsating white dwarf stars.  The exercise provides an
introduction to CCD observing, a framework for teaching concepts in
numerical data analysis and Fourier techniques, and connections with
the physics of white dwarf stars.

(Submitted to the American Journal of Physics.)
\end{abstract}

\maketitle

\section{Introduction}

Astronomers are limited to simply observing the physical universe, and
one of the key tools is recording the brightness of some object of
interest as a function of time.

This technique of \emph{time-series photometry} has yielded many
important results over past decades, particularly in the area of optical
astronomy. These successes include the study of eclipsing binary
stars, pulsating stars and other types of stellar variability.

Astronomy research at Victoria University of Wellington (VUW) over the
past few decades has involved the study of pulsating white dwarfs
using time-series photometry techniques.  A succession of
photomultiplier-based instruments\cite{3chanphot} for obtaining light
curve data has evolved into an efficient and fast CCD (Charge Coupled
Device) camera system that we call \emph{Puoko-nui}\cite{puokonui}.
There are in fact two flavours of the instrument: \emph{Puoko-nui
  South}, based at Mt John University Observatory in New Zealand and
\emph{Puoko-nui North} based at the University of Texas in Austin,
USA.

As part of the development of this CCD photometer, a light box with
controllable light emitting diodes (LEDs) as artificial stars was
constructed for bench testing the instrument.  A very useful addition
was the provision of microprocessor control of the LED intensities,
which enabled the creation via software of a variety of simulation
light curves.

It is basically a modified version of this testing apparatus that we
describe here.  For the laboratory instrumentation, we have replaced
the research grade frame transfer CCD system with a much cheaper
commercial CCD camera from the SBIG\cite{sbig} corporation (SBIG
ST-402ME; the same variety of CCD that we use for autoguiding the
telescope with \emph{Puoko-nui South} attached), and the custom LED
controller was redesigned using an Arduino\cite{arduino} board and
through-hole electronic components in order to simplify reproduction
possibilities.

The particular class of pulsating star of interest here, and the
primary motivation for our laboratory simulation experiment, is that
of the pulsating white dwarf stars\cite{winget}.  These stars exhibit
luminosity variations with timescales in the range of $\sim\,10^{2}$\,s
to $\sim\,10^{3}$\,s.  Monitoring these variations and extracting
frequency information from the resulting light curve enables the
observer to identify pulsation (normal) modes of the star, and thereby
provide information about its otherwise hidden internal structure.
The technique is called \emph{asteroseismology}\cite{winget} due to
its similarity to seismic studies of the Earth's internal structure.

There are a number reasons why simulating the photometric study of the
pulsating white dwarf (WD) stars yields a quality laboratory
experiment.  First, isolated pulsating WDs exhibit extremely coherent
frequencies, and the analysis of the resulting light curves provides
an excellent introduction to the usefulness of Fourier techniques in
such scenarios.  Second, the hundreds-of-seconds periodicities
observed in real stars means that the laboratory experiment can be
operated with timescales matching or nearly matching an actual
observing run at a telescope.  Third, simulating the light output
variations from a real astrophysical object, as distinct from simply
creating a variable signal to analyze, yields a less artificial
experiment and permits natural connections to the underlying physics
of the object.  And, fourth, use of a modern CCD camera provides a
direct introduction to a key instrument that dominates the current
world of optical observational astronomy.

\begin{figure*}
\centering
\includegraphics[width=\textwidth]{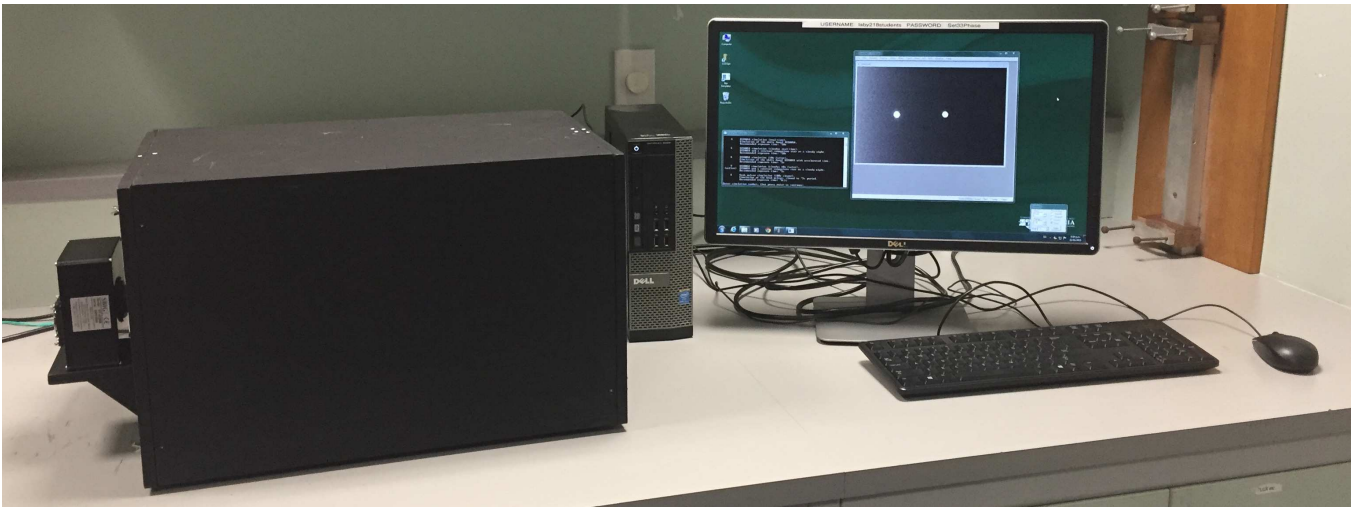}
\caption{A picture of the experimental apparatus: the light box, star
  simulator (hidden), CCD camera, and acquisition computer.}
\label{figure:experimentalhardware}
\end{figure*}

In the following sections we describe in detail the experimental
apparatus and the analytical tools required to extract the
periodicities that are present in the detected time-series data.  We
also provide a few insights into the physics of pulsating WDs.

\begin{figure}[h!]
\centering
\includegraphics[width=\columnwidth]{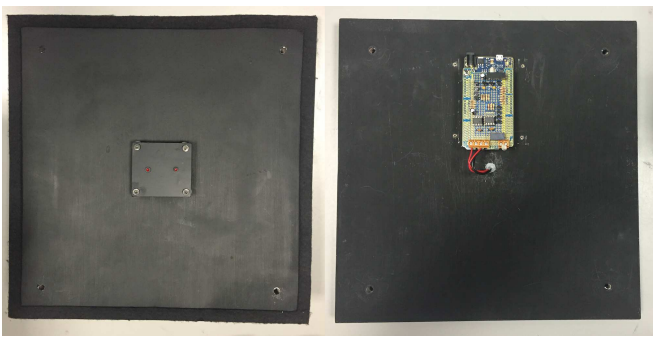}
\caption{Close-up photos of one end plate of the light box show the LEDs and 
the star simulator unit that drives their intensity.}
\label{figure:starsimphoto}
\end{figure}

\section{Experimental Apparatus}

The experimental apparatus (pictured in
Fig.\ \ref{figure:experimentalhardware}) consists of a light-tight box
with the SBIG CCD camera and a small lens at one end imaging a pair of
LEDs attached to a plate on the inside of the box at the other end.
This configuration allows the CCD to be removed for general
inspection, and to adjust the relative position of the lens in order
to optimize the focus.

A key part of the apparatus is the `star simulator' unit (pictured in
Fig.\ \ref{figure:starsimphoto}) that drives the LEDs with simulated
intensity profiles corresponding to the behaviour of chosen
astronomical objects.  It is attached to a panel of the box at the
opposite end to the CCD and connects to the LEDs which are on the
inside.

\begin{figure*}
\centering
\includegraphics[width=\textwidth]{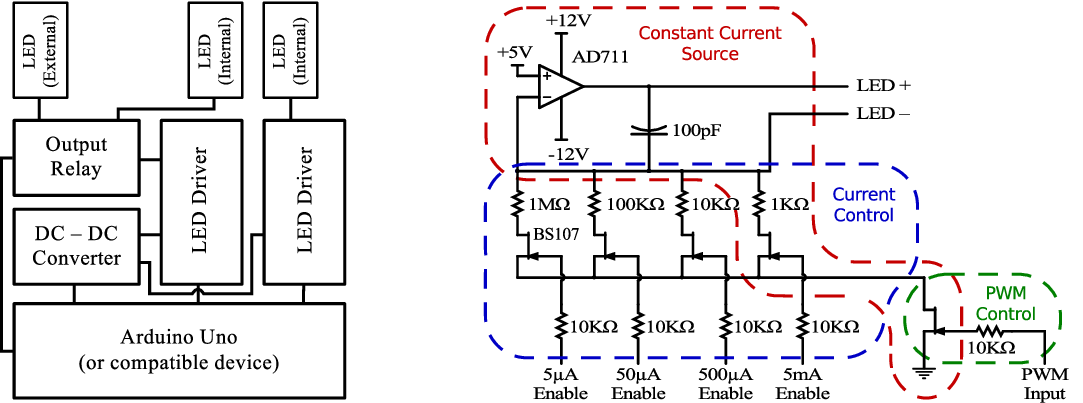}
\caption{Block diagrams depicting the star simulator hardware.  The
  left-hand diagram shows the functional components of the simulator.
  The right-hand diagram shows the LED driver circuit, which includes
  pulse-width modulated constant-current
  sources.}\label{figure:starsimblockdiagram}
\end{figure*}

The core of the simulator electronics is an Arduino\cite{arduino}
microcontroller, with a custom daughter-board (made from an
off-the-shelf prototyping `shield' and through-hole electronic
components) that holds the LED drivers.  A DC-DC converter outputs
$\pm$12V to power the drivers, and an output relay is included such
that one of the output channels can be redirected to an external LED,
which can be viewed directly by eye outside the box.
Fig.\ \ref{figure:starsimblockdiagram} gives a schematic overview of
the star simulator hardware.

Each of the two LED drivers are built around an AD711 operation
amplifier (op-amp) configured as a constant current source.  In this
configuration, the output current is determined by the resistance
between the anode of the LED and ground.  The output driver provides
four software-selectable currents (5\,$\mu$A, 50\,$\mu$A, 500\,$\mu$A,
5\,mA) by connecting four parallel resistors (with different values)
in series with individual BS107 field-effect transistors.  These
transistors are connected to digital output pins on the Arduino, and
act as software-controlled switches that can be switched set the
effective resistance to ground.  All four paths run through a fifth
transistor, which is connected to one of the pulse width modulation
(PWM) outputs on the Arduino.  This transistor is used to rapidly
switch the LED on and off (using a 10-bit timer at 250\,Hz) to provide
1024 fine-grained intensity levels between zero and the maximum
intensity set by the current source.  The combination of current and
PWM control provide the large dynamic range that is required for our
simulations.

The software programming (firmware) inside the Arduino is relatively
straightforward.  When the unit is powered on it reads the desired
simulation parameters from read-only memory, and configures the
current selector transistors and the PWM timers.  The unit then enters
a mode where it will evaluate and update the LED intensities every
16\,ms and checks for new commands from the USB connection to the
control PC.

We have developed four types of simulation profiles for our
experiment: (a) a constant intensity; (b) a sum of multiple sinusoids
(with arbitrary amplitude, frequency and phase); (c) a sum of multiple
gaussians (with fixed repeating period, but arbitrary amplitude,
width, and offset within a cycle); (d) a repeating linear ramp (with
configurable period).  These profiles provide the building blocks to
define a range of simulations (such as the examples described in
Section \ref{section:example}) and apparatus tests.

In addition to these profile types, a cloud simulation can be enabled
on a per-profile basis to attenuate both output channels in a
semi-random manner.  A simple random number generator (based on the
relative drift of two on-board oscillators) seeds control point values
for a spline interpolation algorithm, which results in a smoothly
varying attenuation as a factor of time.  Additional per-point noise
could be included to provide a more realistic simulation, but we omit
this for the sake of experimental simplicity.

The profile parameters for each simulation are defined in the
firmware, and uploaded to the unit via its USB connection.  The active
profile can be changed at any time by running a simple configuration
tool on the acquisition computer.

\section{Data acquisition and analysis}

The main steps required in the experiment are (a) acquisition of a
sequence of CCD frames imaging the illuminated LEDs, (b) extraction of
the LED intensities from each of the frames. (c) conversion of the LED
intensities to form an appropriate time series for Fourier analysis,
and (d) performing the Fourier analysis on various time series light
curves and interpreting the results.  We discuss each of these steps
in turn.

\subsection{CCD frame acquisition}

After selection of one of the simulation light curve options, the
first step is to acquire a sequence of $N$ CCD frames using a chosen
sampling cycle time of $t_{\rm R}$\,s.  Each frame will have some
exposure length $\Delta t$\,s, which will typically be shorter than
$t_{\rm R}$\,s by some amount due to the CCD readout time and possibly
other hardware constraints.  With the SBIG CCD (ST-402ME) we use in
our laboratory experiment, the minimum value for $t_{\rm R}$ is about
1\,s, but the exposure times can be much shorter as the device uses a
mechanical shutter (which is standard) to block the light input during
the readout phase.

Clearly some thought needs to be put into deciding on the these values
for a chosen simulation run such that the Nyquist criterion (a minimum
of two samples for the shortest period) is met for the highest
frequency in the light curve, and the run length $Nt_{\rm R}$\,s is
adequate to provided sufficient frequency resolution in the Fourier
analysis.

In order to provide uniformity, a conventional CCD chip will measure
the stored charge (formally electron-hole pairs in the Si
semiconductor substrate) in each pixel serially, using a single
amplifier and ADC (Analogue to Digital Converter).  This conversion
can take quite some time (seconds or more), especially when operated
at slow speeds in order to reduce measurement noise; consequently,
exposure intervals even in the many seconds domain become rather
inefficient and one can end up closing the telescope to scarce photons
for a significant fraction of an observing run.

The \emph{Puoko-nui}\cite{puokonui} photometers employ frame transfer
CCDs, which have adjacent masked pixel arrays of the same size that
act as storage regions.  The integrated charges in the exposure frames
can be rapidly shifted to the storage frames (in milliseconds) and the
frames can then be digitized and read out in parallel with the next
exposure.  Thus the CCD photometers can be operated without mechanical
shutters, and the deadtime losses become negligible.

The laboratory simulation experiment of course doesn't suffer from a
deficit of photons so the standard CCD is perfectly adequate for the
task in hand.  The SBIG CCD has a fast ($\sim$ms) readout time, and so
the readout dead time is not of practical concern for this experiment.
Thinking in terms of LED brightness variations with periods of
$\sim\,10^{2}$\,s to match the real white dwarf variations, exposure
times of $\sim\,10$\,s or less are appropriate and provide sample rates
significantly higher than the minimum Nyquist rate corresponding to
the highest frequency variation in the light curves.

For their use in astronomical observing, CCDs need to be cooled below
ambient temperatures in order to minimize the thermal creation of
electron-hole pairs that contribute to the background.  This can now
be mostly achieved thermoelectrically; for example when observing, we
cool our \emph{Puoko-nui South} primary CCD to $-50\,$C.  For the laboratory
experiment, with control over the brightness of the LEDs, this can be
avoided, but in the interests of alerting students to the issue it is
sensible to implement cooling.  Our SBIG CCD can be thermoelectrically
cooled and we typically operate it at $-5$\,C.

Another factor which shouldn't remain hidden from students is the
issue of \emph{dark} frames.  These are exposures with the shutter
closed and employing the same time interval as the observation frames:
they are subtracted from the latter to produce the \emph{science}
frames.  With our SBIG CCD, the manufacturer-provided acquisition
software includes options to automate this procedure.  In real
astronomical observing, after dark frame subtraction, the remaining
background is dominated by the sky brightness, which will vary with
the conditions; for the experiment in a light-tight box, this should
in theory be zero.  In practice, however, a finite charge is added to
each pixel before digitization in order to ensure better linearity.
Zero time exposures to measure these values are known as \emph{bias}
frames in the CCD literature.  The dark subtraction also removes this
bias charge.  This fact should at least be mentioned in any notes in
order to satisfy questions from any inquisitive students.

It is perhaps relevant to point out that when using our frame transfer
CCD in shutterless mode, we need to accumulate a separate library of
suitable CCD dark frames outside an observing run.

Although it is not a required step in the laboratory experiment, it
would be appropriate to at least mention the part played by ``flat
fields'' in CCD astronomical photometry.  Invariably because of the
telescope optics, an external uniformly bright field does not produce
a uniform pixel count across the CCD.  This can be accounted for by
observing a physical flat field in high precision and scaling the
science frame pixel counts accordingly.  In practice, a flat field is
obtained in astronomy by observing the bright clear sky at dusk or
dawn (and removing any stellar images) or a uniformly illuminated
board in the telescope dome. (And we will simply mention the debate
concerning which is better).

Finally, the brightest intensity of the LEDs should be monitored, and
the exposure time should be chosen such that the image intensity will not
saturate.  The ADCs used to digitize the pixel charges have a finite
number of digital bits.  In the case of our SBIG CCD, a 16 bit ADC is
used yielding a maximum count of $2^{16} - 1 = 65\,535$ ADU (analog to
digital units).  In practice it is best to maintain the signal of
interest at a level below this maximum value.

\begin{figure}[!ht]
\centering
\includegraphics[width=\columnwidth]{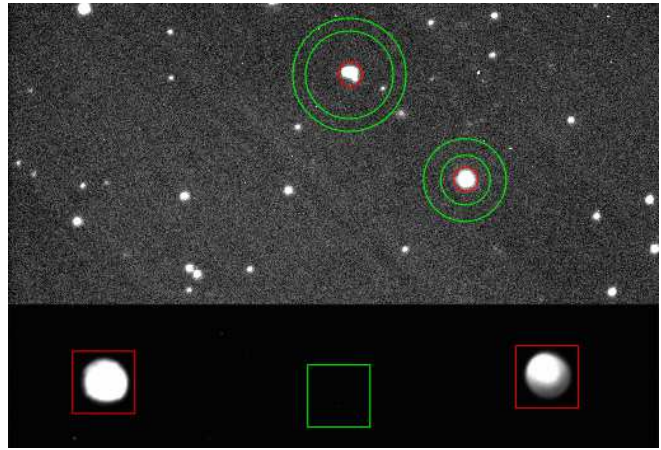}
\caption{A composite picture comparing the two applications of the
  synthetic aperture technique discussed in the text in order to
  extract intensities from CCD frames.  The upper section of the
  figure corresponds to a real observing frame (of the object QU Tel)
  and depicts the circular apertures, while the lower section is an
  image of two light box LEDs and illustrates the simpler square
  aperture method.}
\label{figure:reduction}
\end{figure}

The number of exposures collected for the time series should ensure
that a reasonable number of complete cycles of the longest period are
covered and understanding the significance of this can also be part of
the experiment.

\subsection{Extracting LED intensities}

For the real observing situation with \emph{Puoko-nui} we have written
code that employs circular synthetic apertures to extract stellar
intensities from each exposure frame.  The software package,
\texttt{tsreduce}\cite{puokonui}, uses two concentric apertures
centered on each stellar image of interest to obtain the stellar
intensity above the sky background, and the developing light curves
are presented in real time as the observing proceeds.  The top section
of Fig.\ \ref{figure:reduction} illustrates the technique using a real
observational science frame of the pulsating white dwarf QU
Tel\cite{xcov15} (also known as EC\,20058$-$5234).  It was obtained as
part of an observing campaign in 2011 with the \emph{Puoko-nui South}
photometer attached to the Mt John one metre telescope in New Zealand.
The top encircled image corresponds to the pulsating white dwarf,
while the other object acts as a comparison.

For the simulation experiment the optics are much simpler and one need
not be unduly concerned with the circular symmetry of telescope optics
and the resulting point spread function (including the effects of
astronomical ``seeing'' at the observing site).  Instead it is both
simpler and adequate to employ synthetic square apertures to determine
the integrated LED flux and the background value (refer to the bottom
section of Fig.\ \ref{figure:reduction}).  A close to optimum estimate
of the LED intensity per frame is obtained by subtracting the summed
pixels in the background aperture from the summed pixels in the LED
enclosed aperture.

This approach is straightforward to implement in a computing environment
such as MATLAB or Python, and if deemed appropriate the students themselves
can develop their own extraction code as part of the experimental method.

It is relevant to point out that the synthetic aperture technique
(square or circular) for extracting intensities is only really
suitable for sparse stellar fields; with the aim of making connections
where appropriate with real observing, we simply mention here the
techniques of ``point spread function'' fitting and ``difference
imaging'' for conducting photometry in more crowded fields.  We refer
the interested reader to the extensive literature for explanations of
these techniques.

\subsection{Creation of the time series}

The information of interest in the light curves is encoded in the
variations about the mean level. The actual detected intensity is
rarely of interest in the world of astronomy and one is almost always
required to account for the effects of flux variations created by
cloud interference and changing airmass.  One approach to making these
corrections is to compute the normalized ratio of the variable star
intensity to a nearby (constant) comparison star.  This situation can
be simulated if required in the experiment.

The appropriate form for the time-series data prior to performing a
Fourier analysis is to compute modulation values: percentage variation
in intensity units are an obvious scale to use, but millimodulation
intensity (mmi) units are often used in the white dwarf pulsation
literature as they are suitable for the typical $\sim 1$\% intensity
variations of these stars: 10\,mmi corresponds to a 1\% modulation in
the time domain.  Units consistent with these in Fourier amplitude
space (see below) are millimodulation amplitude (mma) values.
Furthermore, for periodicities in the $\sim10^{2}$\,s region,
$10^{-3}$\,Hz or mHz are suitable for frequency values.  We will use
these units in this paper.

\subsection{Fourier analysis of the time-series data}

Starting with the time series data featuring modulation intensities
versus time, this step involves ``simply'' computing a discrete
Fourier Transform (DFT) of the data.  However, given the simulation
data to be analyzed can include multiple frequencies (with close
spacings) and has high frequency stability and coherence (that match the
observed white dwarf pulsations), there are a number of features and
techniques that can be exploited that will give students good insights
into the use of Fourier analysis.  We will simply summarize the main
features here and allude to their use in the world of pulsating white
dwarf asteroseismic research.

First, it is important to note that the analysis will be using a
discrete version of the Fourier Transform, and to be aware of its
properties and limitations.  The phase information is of
no interest in WD asteroseismic research, so computing the sinusoidal
power (the power spectrum) will yield the amplitude of the
periodicities; and, as it is best to think in terms of amplitudes
rather than power (see below) we require the square root of the power
spectrum, which we will call the Fourier amplitude spectrum or simply
DFT henceforth.

No doubt in most circumstances the DFT will be obtained by a ``black
box'' function (by using MATLAB routines for example), but it is
important to make clear in student minds the difference between a DFT
and the FFT (Fast Fourier Transform).  The FFT is nothing more than an
efficient way of computing a DFT --- with definite limitations.  For
an FFT algorithm the samples must be evenly spaced, and in order to
achieve maximum efficiency the number of samples needs to be a power
of 2 (which is often achieved in practice by zero padding).
Furthermore, the number of points in the frequency domain is dictated
by the number of time domain samples; this can be a limitation
when inspecting a FFT of real data, particularly with noise included.

Given the speed of modern computers we have found in practice that it
is much more useful to use the most simple (but inefficient) DFT algorithm,
with direct control over the number and spacing of the frequency domain
estimates.  This is especially so and essential when combining different
data sets that have different sampling intervals and times, as sometimes
occurs in multi-site observing campaigns run by groups such as the Whole
Earth Telescope (WET)\cite{wet,xcov15}.

Second, we need to introduce the concept of the DFT window and, for
real life observing campaigns, generalize the concept somewhat.  In
the general Fourier literature the DFT standard window function is the
sinc ($\sin x/x$) function (actually sinc$^{2}$ for power), which is
the Fourier transform of a rectangular ``top-hat'' function that
defines the finite length of the time domain sample. The DFT of the
actual data is then a convolution of this sinc function and the
Fourier transform of the theoretically infinite sinusoids in the data.
In some applications, tapering the ``top-hat'' ends in order to reduce
the amplitude of the wiggles in the sinc function (at the expense of
broadening the central peak) is considered effective.  This is called
\emph{apodisation} in the literature.

However, in the world of pulsating white dwarfs in particular, with
the occurrence of unavoidable data gaps due to such things as cloud
interference and the rising of the Sun, what is needed is a direct
transformation of the signal turning on and off (often multiple times)
into the Fourier domain.  This is achieved by computing the DFT of a
single noise free sinusoid with a suitable period calculated for each
sample time for the real data.  This function can be compared with
various peaks in the DFT of the real data in order to see if they are
consistent with only one frequency.  Even if the experiment is limited
to investigating the noise-free simulation light curves, it is
instructive to compute the window function when examining light curves
with multiple gaps and closely spaced frequencies.

Finally, in the interests of completeness and making connections with
real observational data, it is relevant to discuss the impact of noise
on the frequency identification process.  As part of generating the
light curves, observational uncertainties or noise can easily be added
using for example an algorithmic (pseudo-)random number generator.
The issue of identifying as real low level peaks in the DFT computed
from an observational light curve is an important part of the analysis
in observational astronomy.

A technique that has proved to be useful, especially in the case of
low amplitude periodicities that are close to large amplitude ones, is
referred to as \emph{prewhitening}\cite{xcov15}.  One identifies the
periods of the larger amplitude peaks in the DFT and then uses these
to compute a synthetic light curve.  This synthetic light curve is then
subtracted from the real time-series data and another DFT calculated
for the now ``prewhitened'' data.  This new DFT should provide a
clearer indication of the reality of any remaining peaks above any
nearby noise peaks.

If the periodicities $T_{i}$ of the peaks being subtracted are known
accurately enough, they can be fixed in a straight forward
\emph{linear} least squares procedure by fitting values for $A_{i}$
and $B_{i}$ in functions of the following form

\begin{equation*}
 f(t) = A_{i}\sin\left[\frac{2\pi t}{T_{i}}\right] 
      + B_{i}\cos\left[\frac{2\pi t}{T_{i}}\right]
      = C_{i}\sin\left[\frac{2\pi t}{T_{i}} + \phi_{i}\right]
\end{equation*}

Alternatively, the $T_{i}$ values can also be optimized using
\emph{nonlinear} least squares techniques, which require iterative
procedures.  In this case, it would more convenient to also optimize
the parameters $C_{i}$ and $\phi_{i}$ in the above functions.

\begin{figure}[!ht]
\centering
\includegraphics[width=\columnwidth]{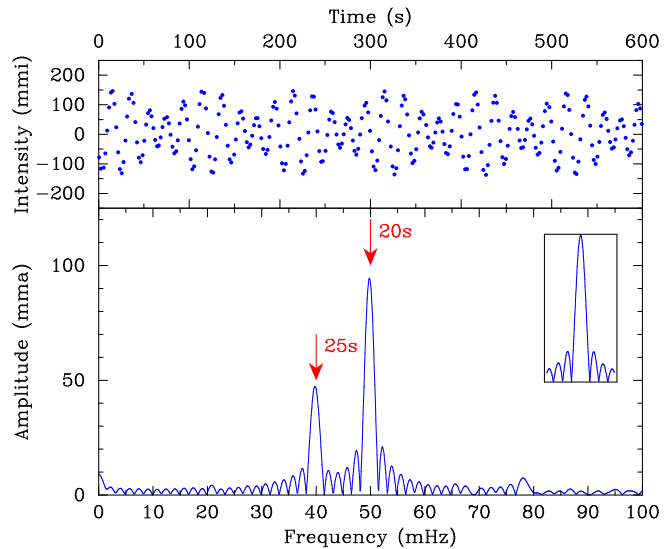}
\caption{Modulation time series data (top panel) and the corresponding
  amplitude DFT (bottom panel) for a 600\,s run using a sampling rate
  if 1\,s.  The LED intensity has been modulated by two frequencies
  with periods of 20\,s and 25\,s and different amplitudes, which leads
  to the visible beating.  The sample window function is shown in the insert.}
\label{figure:beating}
\end{figure}

It is instructive to compare the amplitudes $C_{i}$ of the least
squares derived sinusoids with those in the DFT amplitude spectrum.
Of course something would be radically wrong if they were not very
similar, but making the comparison reinforces the basic fact that the
least squares estimates and the DFT amplitudes \emph{should} be very
similar even though they are derived from apparently
different procedures.  With this comparison in mind, it is useful to
view the DFT as a set of cross correlations of the time-series data
with a sequence of sinusoids with different frequencies.

When deciding whether a peak in a DFT corresponds to a real
periodicity in the light curve, it is important to have a statistical
model for the time domain noise and its effect on the uncertainties in
the DFT.  With the speed of modern computers, simulation procedures
can make this task largely model-free.  We are attracted to a
simulation implementation\cite{xcov15} of the false alarm probability
(FAP) concept\cite{scargle} for characterizing uncorrelated noise
levels in DFTs.  This method is intuitively straight forward and
essentially free of complicated statistical theory.  It is certainly
preferable to simply quoting some formula for the DFT noise level.

One first \emph{prewhitens} (removes) the main periodicities from the
light curve, leaving residuals that characterize the time-domain
noise level.  The time stamps for each of the sample values are then
randomly shuffled, thus destroying any remaining coherent signals in
the data, but leaving the random noise level intact.  A DFT of the
time-shuffled data is calculated and the height of the highest peak in
the region of interest is noted.  This procedure is repeated N times
and a record kept of each highest peak, resulting in an ensemble
of highest peaks.  One then asserts that there is a one in $N$ chance
of the highest ensemble peak occurring through a noise conspiracy in
the data.  For $N=1000$ say the chance or FAP is 1 in 1000.  Thus, if
a peak is visible in the DFT at this level, there is 0.1\% chance
of it simply being noise. It is instructive to construct a histogram
of the ensemble of highest peaks obtained in the simulation --- some examples
can be found in the references\cite{xcov15}.

The implementation and effectiveness (along with some explanations) of
the above techniques can be found in a papers describing observational
programs on pulsating white dwarfs\cite{xcov15,ec04207}.  In
particular, an implementation of a FAP simulation procedure should really
make use of the much more efficient FFT algorithm given the requirement
to compute a large sample of DFTs.

\begin{figure*}
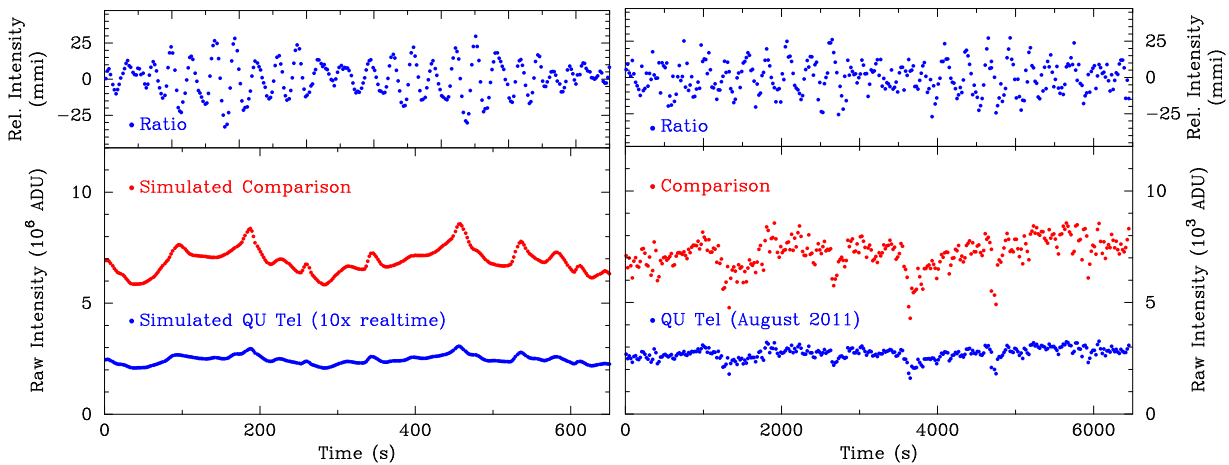

\centering
\includegraphics[width=0.45\textwidth]{ecsim.ps}
\includegraphics[width=0.45\textwidth]{ecreal.ps}
\caption{The low level white dwarf pulsations can be disguised or
  completely hidden by cloud variations.  The simulated cloud effect
  (left panel) features a smooth intensity change, but avoids the
  large per-point noise that is often introduced when observing
  through real cloud (right panel).  In both cases, the WD pulsations
  can be recovered by dividing the (absolute) intensity of the target
  star by a constant comparison star: the cloud affects both
  measurements equally (however note that there are second-order
  effects that reduce this in real observations), and this
  multiplicative factor cancels in the division.
}\label{figure:ec20058cloud}
\end{figure*}
 
\section{Some Example Procedures}\label{section:example}

In this section we provide three example cases with different
light curves.

The first one involves sampling a light curve that features beating
between two sinusoidal signals.  This provides a good starting example
as it provides a relatively simple signal while the
students are familiarizing themselves with the general operation of
the camera and the data reduction procedures.  It also illustrates the
main features associated with sampling a signal with two adjacent
sinusoidal signals over a finite interval.  Fig.\ \ref{figure:beating}
depicts both the set of modulated time series samples covering an
interval of 600\,s and the resulting DFT.  The DFT window function for
this sample is also shown in the insert.

The second example employs a light curve generated using multiple
frequencies corresponding to those observed in the pulsating white
dwarf QU Tel\cite{xcov15} (also known as EC\,20058$-$5234).  Cloud
simulation (affecting both LED intensities) is also added to the mix
and the correct light curve variations recovered by computing intensity
ratios after background estimates have been subtracted (refer to
Fig.\ref{figure:ec20058cloud}).

\begin{figure}[!ht]
\centering
\includegraphics[width=0.9\columnwidth]{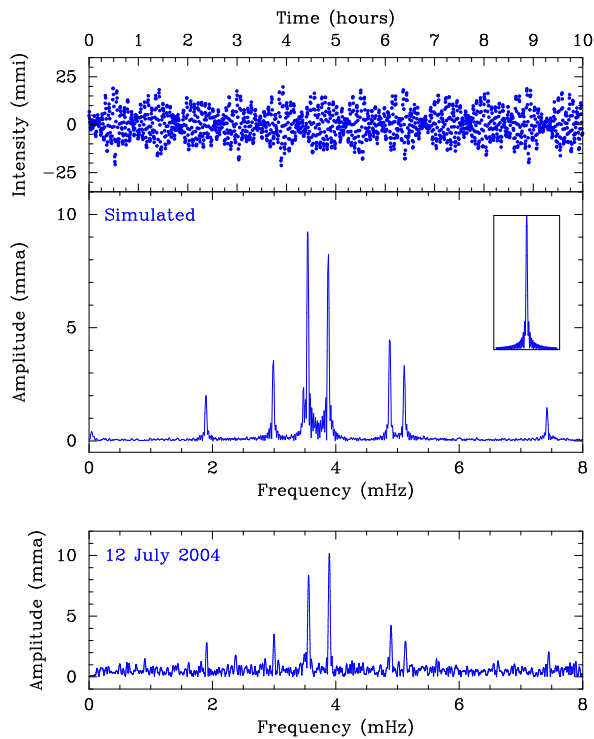}
\caption{A 10\,hour collection run (top) on a ``cloudless'' real-time
  simulation of QU Tel provides a very realistic result when compared
  with real observational data (bottom).  The insert in the upper
  figure shows the DFT window for the siumulation.  This simulation
  configuration would be suitable for a longer-form project where the
  students are able to set up a data collection run overnight.  A
  shorter collection period can be used at the expense of a poorer
  frequency domain resolution.  Alternatively, the frequency and
  intensity can be increased by 10$\times$ to accelerate the data
  collection time to only 1 hour.}
\label{figure:ec20058dft}
\end{figure}

Fig.\ \ref{figure:ec20058dft} compares a 10\,h run using the QU Tel
simulated light curve with a real $\sim$10\,h 2004 observing run on
this pulsating white dwarf carried out at Mt John Observatory in New
Zealand \cite{xcov15}.  Included in the figure is a comparison of the
two DFTs.

The final example illustrates the principles of signal folding, and
provides an impressive example of the ability to reconstruct periodic
signals from sparsely sampled data when there is a high level of signal
coherency.  The simulation profile is based on optical observations of
the pulsar (PSR\,B1919+21) in the Crab nebula that were acquired in
2013 using the \emph{Puoko-nui North} photometer\cite{puokonui}.

A pulsar is a rotating neutron star that emits two beams of radiation
such that if one of the beams intersects the Earth it is detected
(primarily) in the radio frequency domain as an extremely stable
periodic signal.  The Crab pulsar is one of a handful that have also
been detected optically.  The first Crab pulsar optical detections
were obtained via photometry synchronized to the known radio ephemeris
\cite{crab_cocke,crab_nather}, but one can also achieve the detection
using signal folding as shown in this simulation example.

Because the SBIG camera has a maximum frame rate of 1 frame per
second, the 33\,ms period of the simulation has been increased by a
factor of 100\,$\times$ to a more manageable $\sim3.3\,s$; this yields
roughly three frames sampled per pulsation cycle.

\begin{figure}[!ht]
\centering
\includegraphics[width=\columnwidth]{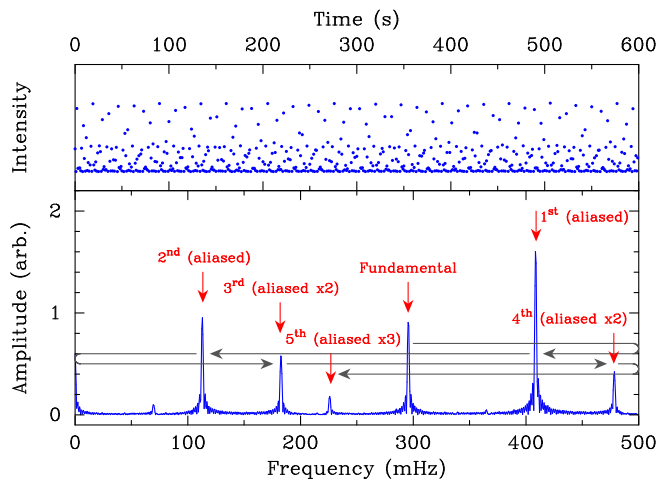}
\caption{The time series data (upper panel) obtained as part of the
  pulsar example and its DFT (lower panel).  This example illustrates
  the effect of DFT aliasing in undersampled signals and the
  utility of data folding at a known period (refer to text).}
\label{figure:crabdft}
\end{figure}

Figure \ref{figure:crabdft} shows the raw time series acquired in a
10\,minute collection run and the DFT of this data.  The pulsation
profile is non-sinusoidal (it is modelled using two gaussians), and so
the main features in the frequency domain are the fundamental
pulsation period and its harmonics.  The shape of the pulsation
profile in the time domain are encoded in the relative amplitude and
phase of the harmonics in the frequency domain.

The DFT plot shows another important aspect of Fourier analysis:
frequencies that are greater than the Nyquist frequency (500\,mHz for
these 1\,s samples) are `aliased' back inside the DFT range by
reflection around either the Nyquist frequency (at the high end) or
zero (at the low).  It can be difficult to distinguish aliases from
true signals, but in this case we can uniquely identify the
fundamental frequency based on the known time-domain period (which can
be timed manually using the external LED on the lightbox), and it is
then a straightforward exercise to identify the remaining peaks by
taking integer multiples of the fundamental frequency and applying the
appropriate reflections.

\begin{figure}[!ht]
\centering
\includegraphics[width=\columnwidth,angle=-90]{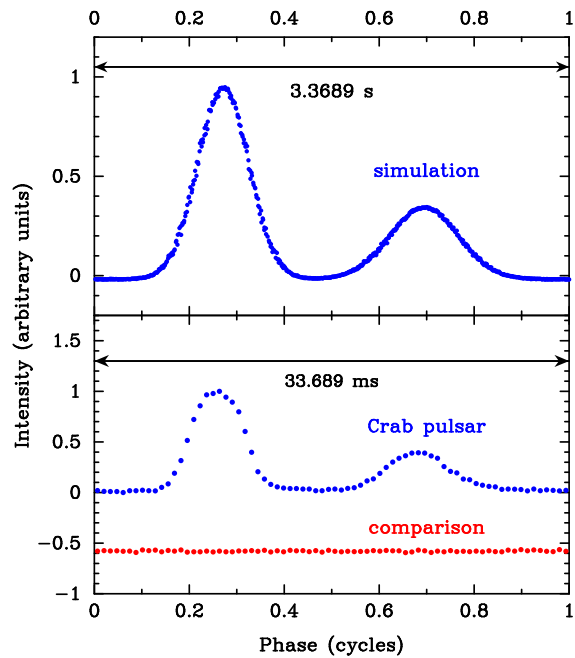}
\caption{The simulated pulsar time series data folded at the known
  period (upper plot) compared with the actual Crab pulsar folded data
  (and a nearby comparison) obtained in 2013 using the \emph{Puoko-nui
    North} photometer. The Crab and comparison data have been binned
  (see reference\cite{puokonui}).}
\label{figure:crabfold}
\end{figure}

The precise period of the pulsar signal can be identified from the DFT
plot, and this period is then used to `fold' the signal over itself in
order to create a detailed form for a single pulsation cycle.  This
can be represented mathematically using the equation

\begin{equation*}
	\phi_i = \frac{t_i}{T} - \left\lfloor\frac{t_i}{T}\right\rfloor\,,
\end{equation*}

where $\left\lfloor \ \right\rfloor$ denotes the integer division
operator (divide and discard the remainder).  In other words, we are
converting the time-based signal to a phase (which may be itself
measured as a time, angle, or fractional number).  This is shown in
Figure \ref{figure:crabfold}, which compares the simulated pulsar
photometry against our real observational data.

\section{Discussion}

We have outlined a range of topics connected with the observation of
pulsating white dwarf stars that can be explored using our simulation
apparatus.  Depending on the course level that the experiment might be
offered in, one would want to make a selection of the topics
discussed.  During the early development of this experiment we used it
very successfully as the basis of a one semester project in
2013 in our 400-level Honours course in physics, and in 2014 it was
run as a 200-level experiment.

Ideally, a laboratory experiment using the simulator should endeavor
to draw as many links as possible to the real world of astronomical
observing and the physics of the target objects.  The pulsating white
dwarfs are certainly a suitable subject for this task.

When white dwarfs pulsate (at several well-defined surface temperature
regimes) they are usually multiperiodic and exhibit very stable
frequencies.  This is largely a consequence of their compact very
slowly evolving structure -- about the mass of the Sun compressed into
a sphere about the size of the Earth.  They are prevented from
gravitational collapse by the temperature-independent degeneracy
pressure resulting from the essentially separate electron gas in the
plasma being gravitationally confined at very high number densities.

The observed luminosity variations result from nonradial \emph{gravity
  mode} pulsations --- the more common and intuitively simpler radial
pulsations due to pressure variations (seen in many other stellar
types) have never been observed in white dwarfs.  The underlying
driving mechanism for stellar pulsation is the interaction of the
escaping core radiation with a partially ionized chemical element
(usually H or He) in a layer near the star's surface.  This mechanism
was first postulated by Eddington\cite{eddington} early last century.

For white dwarfs, the very high surface gravities clearly inhibit
radial material motion and hence limit any consequent luminosity
variations.  In addition, the large bulk modulus (high ``stiffness'')
of the material leads to high sound speeds and hence to expected
periodicities of seconds or less for the pressure modes.  These
oscillations, if they exist, are likely to be very difficult to
detect\cite{silvotti11,kilkenny14}.  

The gravity modes, however, are the result of material density
gradients in the local gravitational field and feature much longer
periods: the so-called Brunt-V\"{a}is\"{a}l\"{a}
frequency\cite{winget} determines these periods and its calculated
values are consistent with the hundreds of seconds luminosity
variations seen in the pulsating white dwarfs.  A simplified and
instructive physical picture for the phenomenon is to envisage the
oscillations of an object floating at the interface between two fluids
of different density (eg water and air).

However, in white dwarfs the gravity modes involve largely nonradial
material motion and a rich array of possible normal modes.  It is
interesting to note that the gravity mode pulsations in white dwarfs
were discovered\cite{landolt} before a theoretical model was proposed.

A final note on the asteroseismic study of white dwarfs is worth
making and should be at least mentioned in any experimental notes.
The detected frequencies in a pulsating white dwarf are used to infer
normal modes of the star and these are used to guide and constrain
theoretical models of the structure.  Including some of these steps in
an experiment is possible, but it would take us beyond the scope of
this article.

Obviously, the apparatus could also be used to simulate other
observational programs by generating appropriate synthetic light
curves: eclipsing binary stars, planetary transit events, or even
gravitational microlensing events involving extrasolar planets
orbiting stars\cite{ulensing} are interesting possibilities.

From an observer's perspective, simulating the observation of the
short period binary object NY Vir (also known as
PG\,1336$-$018)\cite{pg1336,puokonui} provides an interesting
experience: it has a $\sim$2.5\,h period, a deep primary eclipse, a
smaller secondary eclipse combined with a `reflection effect' and one
component is also a pulsating (subdwarf B) star.  A number of classic
binary star light curve characteristics are therefore in evidence, but
more detailed analysis would require some advanced modelling.

Also, the detection of extrasolar planets via transits is certainly of
current topical interest\cite{kepler} and monitoring and analyzing a
simulation light curve corresponding to a relatively rapid planetary
transit\cite{jha,djsts} could make a useful laboratory experiment.

However, we think that the optimum and least artificial use of the
equipment is to produce and analyse light curves corresponding to
pulsating white dwarfs, develop expertise in Fourier techniques and
also explore at least some of the physics of these exotic yet
ubiquitous astrophysical objects.

\begin{acknowledgments}

We acknowledge financial support from the NZ Marsden Fund and
help from the School technical staff --- Rod Brown in particular.

\end{acknowledgments}

\end{document}